\documentclass[twocolumn]{aastex61}

\usepackage{graphicx}
\received{?}
\revised{?}
\accepted{?}
\submitjournal{ApJL}

\shortauthors{Fu et al.}
\newcommand{\oql}{O$^{7+}$/O$^{6+}$}

\newcommand{\qfe}{$Q_{Fe}$}
\newcommand{\velunit}{km~s$^{-1}$}

\newcommand{\ahe}{$A_{He}$}

\begin{document}

%\title{A case study of the compositional properties of ICME and activities in the source
%        regions: from the perspective of mass supply of the CME}
\title{The high helium abundance and charge states of the interplanetary CME and its
        material source on the Sun}

\correspondingauthor{Hui Fu}
\email{fuhui@sdu.edu.cn}

\author{Hui Fu}
\affil{Shandong Key Laboratory of Optical Astronomy and Solar-Terrestrial Environment, Institute of Space Sciences, Shandong University, Weihai, Shandong, 264209, China}
\affil{STFC-RAL Space, Rutherford Appleton Laboratory, Harwell Campus, Didcot, OX11 0QX, UK}

\author{R.A. Harrison}
\affil{STFC-RAL Space, Rutherford Appleton Laboratory, Harwell Campus, Didcot, OX11 0QX, UK}

\author{J.A. Davies}
\affil{STFC-RAL Space, Rutherford Appleton Laboratory, Harwell Campus, Didcot, OX11 0QX, UK}

\author{LiDong Xia}
\affil{Shandong Key Laboratory of Optical Astronomy and Solar-Terrestrial Environment, Institute of Space Sciences, Shandong University, Weihai, Shandong, 264209, China}

\author{XiaoShuai Zhu}
\affil{Max Planck Institute for Solar System Research, Justus-von-Liebig-Weg 3, 37077, G\"ottingen, Germany}

\author{Bo Li}
\affil{Shandong Key Laboratory of Optical Astronomy and Solar-Terrestrial Environment, Institute of Space Sciences, Shandong University, Weihai, Shandong, 264209, China}

\author{ZhengHua Huang}
\affil{Shandong Key Laboratory of Optical Astronomy and Solar-Terrestrial Environment, Institute of Space Sciences, Shandong University, Weihai, Shandong, 264209, China}

\author{D. Barnes}
\affil{STFC-RAL Space, Rutherford Appleton Laboratory, Harwell Campus, Didcot, OX11 0QX, UK}



\begin{abstract}

%{\bf
%The generation and mass supply of the coronal mass ejections (CMEs) are fundamental issues
%    in solar and space physics.
%Given that the compositional parameters, such as charge states, helium abundance (\ahe)
%    maybe not the same for the plasma generated by different processes and/or originates from %different regions,
%    we examine the connection between the in-situ properties of the CME and activities in the Sun.
Identifying the source of the material
%    often with high charge states and high He abundance (\ahe),
    within coronal mass ejections (CMEs)
    and understanding CME onset mechanisms are fundamental issues
    in solar and space physics.
Parameters relating to plasma composition,
    such as charge states and He abundance (\ahe), may be different
    for plasmas originating from differing processes or regions on the Sun.
Thus,
    it is crucial to examine the relationship between in-situ measurements of CME composition
    and activity on the Sun.
We study the CME that erupted on 2014 September 10, in association with an X1.6 flare,
    by analyzing AIA imaging and IRIS spectroscopic observations and
    its in-situ signatures detected by Wind and ACE.
We find that during the slow expansion and intensity increase of the sigmoid,
    plasma temperatures of 9 MK, and higher, first appear
    at the footpoints of the sigmoid,
    associated with chromospheric brightening.
Then the high-temperature region extends along the sigmoid.
IRIS observations confirm that
    this extension is caused by transportation
    of hot plasma upflow.
Our results show that chromospheric material can be heated to 9 MK,
    and above,
    by chromospheric evaporation
    at the sigmoid footpoints before flare onset.
The heated chromospheric material can transport into
    the sigmoidal structure and supply mass to the CME.
The aforementioned CME mass supply scenario provides a reasonable explanation
    for the detection of high charge states and elevated \ahe\
    in the associated ICME.
The observations also demonstrate that the quasi-steady evolution in the precursor phase
    is dominated by {magnetic} reconnection
    between the rising flux rope and the overlying magnetic field structure.

\end{abstract}




\section{introduction}
\label{sec:intro}

%Coronal mass ejections (CMEs), which were first observed by the Orbiting Solar Observatory (OSO-7)
%    \citep{1973spre.conf..713T},
%     are the coronal signatures of the eruption of huge plasma and magnetic field structures
%     that are ejected from the Sun

%    \citep{2012LRSP....9....3W}.
%Interplanetary coronal mass ejections (ICMEs) are the heliospheric counterparts of the CMEs.
%    \citep{1985JGR....90..163S,2014SoPh..289.3843R,2017SSRv..212.1159M}.
%As the Earth is embedded in a space environment defined by the outflowing solar wind,
%    there are significant periods of time during which the Earth is immersed in outflowing CME %plasma.
The Earth is in coronal mass ejection (CME) outflow for $\sim$10\% of the time
    at solar minimum up to 35\% at solar maximum
    \citep{2003ApJ...592..574C}.
Therefore, the CME generation mechanism and the source of CME material
    are important issues in helio-physics.

%\red{The connection between CMEs and solar flares is crucial for understanding
%    the generation mechanisms for the two phenomena \citep{2003AdSpR..32.2425H}.
%Solar flares \citep{1859MNRAS..20...13C},
%    which are the most powerful energy release process {on the Sun},
%    are strongly associated with CMEs,
%    although this association is not one to one.
%The rate association between CMEs and flares increases with flare class.
%About 10\% (40\%) of C (M) class of flares are accompanied by CMEs
%    \citep{2006ApJ...650L.143Y},
%    and the association rate can reach up to 90\% for X class flares
%    \citep{2007ApJ...665.1428W}.
%}

Studies of the relationship between in-situ parameters,
    such as charge states, helium (He) abundance (\ahe), first ionization potential (FIP) bias
    and activity in the source region
    can promote understanding of mass supply to CMEs
    and the connection between CMEs and flares.
These composition parameters carry information from the near-Sun region
    where CMEs are forming,
    the solar wind is accelerating,
    and the corona is heated
    \citep[e.g.][]{2017SSRv..212.1159M}.
%    \citep[][and the references therein]{2000RvGeo..38..247B, 2004JGRA..109.9104R, %2017SSRv..212.1159M}.
%The number density ratio of {O$^{7+}$} to {O$^{6+}$} (\oql) does not change
%    beyond several solar radii from the Sun.
%    and depends on the plasma density and temperature.
%As the electron density decreases sharply with heliocentric distance,
%    the ions in the solar wind no longer encounter electrons,
%    and the charge states of the ions is ``frozen" in.
Beyond several solar radii,
    the ion charge state,
    such as the number density ratio of {O$^{7+}$} to {O$^{6+}$} (\oql),
    is ``frozen in".
Therefore, the charge states further from the Sun
    represents the temperature of the coronal source region
    \citep{2001JGR...10629231L}.
%    \citep{1983ApJ...275..354O, 1986SoPh..103..347B, 1998GeoRL..25.3465H, 2001JGR...10629231L}.
The average iron (Fe) charge state (\qfe) is {$9^{+}$} to {$11^{+}$}
    in the solar wind \citep{2001JGR...10629231L}.
%    shows that the temperature of the solar wind source regions is lower
%    than the temperature that occurs during a flare.
In contrast, more than 90\% of long-duration higher \qfe\ episodes
    (greater than 12 for more than 6 hours)
    occur inside Interplanetary CMEs (ICMEs) and around half of ICMEs are accompanied by high \qfe\
    \citep{2001JGR...10629231L, 2004JGRA..109.1112L}.
Low FIP elements, such as Mg, Si, and Fe,
    are usually more abundant
    in the solar wind and corona than in the photosphere
    \citep[the FIP bias effect,][]{1992PhyS...46..202F}.
Conversely, the high FIP element helium is usually less abundant
    in the corona and solar wind than in the photosphere.
%\ahe\ is generally lower than 5\% in the solar wind,
%    and changes with solar phases
%    \citep{1974JGR....79.4595O, 2001GeoRL..28.2767A, 2007ApJ...660..901K, 2012ApJ...745..162K,
%   2018MNRAS.478.1884F}.
%Considering the fact that the FIP is highest for the helium and it remains neutral longest,
%    one of the possible explains for the enrichment/depletion of the low FIP elements/helium
%    is that the mechanism which transports the element to the upper atmosphere
%    only interacts with ions \citep{2012ApJ...744..115L, 2015LRSP...12....2L}.
%A possible explanation for
A possible explanation for the enrichment of low FIP elements (i.e. FIP bias effect)
    and depletion
    of helium in the corona/solar wind is that
    the mechanism that transports the element
    to the upper atmosphere is only valid to %leads to interact to
    ions \citep{2015LRSP...12....2L}.
%Therefore, %the abundance of low FIP elements(helium) is
    %high(low) in the upper atmosphere.
%    the low FIP elements(high FIP element helium)
%    are enriched(depleted) in the upper atmosphere.
%    \citep[][and references therein]{2015LRSP...12....2L}.

%The connection between the in-situ signatures of an ICME and activity
%    in its associated source region is highly complicated.
Statistically,
    the charge states, \ahe\ and FIP bias inside ICMEs are higher
    than in the background solar wind
    \citep{2004JGRA..109.9104R, 2017SSRv..212.1159M, 2018SoPh..293..122O}.
However, the variation in these parameters over a single ICME,
    and between different ICMEs,
    can be so significant that they cannot be used for definitive ICME identification
    \citep{2004JGRA..109.9104R}.
%    \citep{1983ApJ...275..354O, 1986SoPh..103..347B, 1998GeoRL..25.3465H, 2001JGR...10629231L}.
In some ICMEs, the source temperatures that are deduced from the charge states
    can exceed 10 MK,
    which can only be achieved by heating by flares \citep{2006SSRv..124..145G}.
Statistical results show that \qfe\ and \oql\ in an ICME
    correlate positively, but only weakly, with flare class
    \citep{2008ApJ...682.1289R, 2013SoPh..284...17G}.
%This correlation is, however, only weak,
%    and in some cases, charge state is not elevated in an ICME associated with a strong flare.
This suggests that plasma that has been heated by flares
    is not always a source of CME material.
%In contrast, in some cases, the charge states are higher inside ICMEs
%    accompanied by weak flares.
\citet{1982JGR....87.7370B} found that ICMEs associated with strong flares,
    on average, have higher \ahe,
    although \citet{2008ApJ...682.1289R} found no clear correlation between \ahe\
    in an ICME and flare intensity.
%However, the statistical results show that the helium abundance of ICMEs is much different case by %case,
%    and it has no clear correlation with the intensity of the flares \citep{2008ApJ...682.1289R}.
The reason for elevated \ahe\ in ICMEs is not clear
    \citep{2017SSRv..212.1159M}.

%The complicated relationship between the abundances of elements within ICMEs
%    and flare intensities
%    should depend on source region activity
%    and the processes that supply CME material.
If flare-heated plasma is injected into CMEs,
    then the accompanying ICMEs should have higher charge states.
Plasma can be heated to temperature of 10 MK and higher by two processes during flares:
    (1) magnetic reconnection in the corona
    \citep{2005ApJ...622.1251L, 2013NatPh...9..489S}
    and (2) chromospheric evaporation
    \citep{2013ApJ...766..127Y, 2013SoPh..288..637T, 2015ApJ...811..139T}.
%Firstly, plasma can be heated directly by directly magnetic reconnection in the coronal
%    \citep{1983ApJ...275..354O, 1986SoPh..103..347B, 1998GeoRL..25.3465H, 2001JGR...10629231L}.
%Secondly, chromospheric evaporation can heat material to temperatures
%    up to tens of millions of degrees
%    \citep{2013ApJ...766..127Y, 2013SoPh..288..637T, 2015ApJ...811..139T, 2018ApJ...856...34T}.
%    \citep{2013ApJ...766..127Y, 2013SoPh..288..637T, 2015ApJ...811..139T, 2017ApJ...841L...9L,
%    2018ApJ...856...34T, 2019ApJ...870..109Z}.
In the latter, electrons are energized by magnetic reconnection in the corona,
    and then propagate downwards where they heat the chromosphere via Coulomb collisions.
    %ward alone magnetic field lines.
%The high-energy non-thermal electrons lose energy
%    via Coulomb collision and impulsively heat the chromospheric material.
    %impulsive heating of chromosphere material.

Generally, it is believed that the hot plasma inside ICMEs can be heated
    directly by magnetic reconnection
%    taking place along the current sheet
    between the flux rope and flare loops during the flaring process
     \citep{2002A&ARv..10..313P, 2015ApJ...808L..15S, 2016ApJS..224...27S}.
%In contrast, the plasma heated by the chromospheric evaporation processes can not
%    supply mass to the ICMEs,
%    as the heated plasma is constrained by the closed flare loops.
%Besides, CMEs leave the solar surface when the chromospheric evaporation processes starts,
%    based on the standard CME and flare model
%    \citep{2002A&ARv..10..313P, 2006SSRv..124..145G, 2015SSRv..194..237L}.
    %standard CSHKP model(Carmichael 1964; Sturrock 1966; Hirayama 1974; Kopp and Pneuman 1976).
%
%
%However, chromospheric evaporation does not only take place during
%    the main phase of the flares,
%    but also occurs before the onset of a flare.
%\citet{2018ApJ...854...26L} observed chromospheric evaporation in the flare precursor phase,
%    although only lasting for several minutes.
%Besides, the internal reconnection,
%    the successive reconnection between the rising flux rope and the overlying field lines,
%    can also produce non-thermal electrons and cause heating at the footpoints of magnetic field %lines
%     \citep[e.g.,][]{2001ApJ...552..833M, 2001JGR...10625227S, 2004ApJ...602.1024S, %2004ApJ...611..545G, 2019ApJ...878...38Y}.
%Internal reconnection may take place during the flare precursor phase
%    (i.e. during the CME initiation phase)
%    as flux ropes generally start to expand slowly before
%    the eruption of CMEs and the onset of flares
%    \citep{2001ApJ...559..452Z, 2019ApJ...871...25W}.
%Therefore, it is worth examining,
%    whether the plasma heated by chromosphere evaporation processes
%    could supply material to CMEs.
%
%
It is not known if the heated chromosphere can supply mass to CMEs.
\ahe\ may not be the same
    for the two different hot %(10 MK and higher)
    plasma sources.
%    as they are heated in different regions
%    and by different mechanisms.
{\ahe\ in a CME should be lower than in the photosphere
    if the plasma is heated by magnetic reconnection directly in the corona,
    as \ahe\ is lower in the corona compared with that in the photosphere
    \citep[][and references therein]{2015LRSP...12....2L}.
In contrast, \ahe\ in plasma heated by chromospheric evaporation
    may be the same as in the photosphere (i.e. higher than in corona)
    if the helium is still neutral prior to being heated.}
%    as the depletion of helium has not works.

%Therefore, it is worth examining
%    whether the plasma heated by chromosphere evaporation processes
%    could supply material to CMEs.

%Thus, it is crucial to analyze the connection between the composition of ICMEs
%    and activity in the source regions in more detail.
{Using the 2014 September 10 CME as a case study,
    we address the question of why \ahe\ can be nearly as high as in the photosphere in some ICMEs,
    and where that material originates.}
In the present study,
    we analyze
    (1) the morphology evolution of a hot channel,
    (2) chromospheric brightening and plasma heating at the sigmoidal footpoints,
    (3) transportation of hot plasma along the sigmoid, and
    (4) the connection between the aforementioned phenomena
    during the precursor phase of the flare
    (corresponding to the CME initiation phase).
The in-situ parameters relating to composition
%    such as charge state and the helium abundance of the associated ICME,
    are also analyzed.

\section{Instruments and Data Analysis} \label{sec:data}
An X1.6 flare took place in NOAA AR (Active Region) 12158,
    with the peak of the
    Geostationary Operational Environmental Satellite (GOES)
    Soft X-Ray (SXR) intensity occurring at about 17:45 UT on 2014 September 10.
The remote-sensing data used to study the solar atmosphere during this event
    are mainly from the Atmospheric Imaging Assembly \citep[AIA;][]{2012SoPh..275...17L}
    and the Helioseismic and Magnetic Imager \citep[HMI;][]{2012SoPh..275..229S}
    onboard the Solar Dynamics Observatory \citep[SDO;][]{2012SoPh..275....3P},
    and the Interface Region Imaging Spectrograph \citep[IRIS;][]{2014SoPh..289.2733D}.
AIA images the solar atmosphere in 7 EUV passbands
    and 3 UV passbands ranging in temperature
    from $6 \times 10^{4}K$ to $2 \times 10^{7}K$.
The spatial resolution of AIA is 1.2 arcsec
    and the temporal resolution is 12s and 24s for EUV and UV passbands, respectively.
HMI provides measurements of the vector magnetic field in the photosphere.
IRIS can obtain spectral data and slit-jaw images simultaneously.
For the former, the spatial resolution is 0.166 arcsec per pixel,
    and the cadence is 9.5s.
IRIS observations were taken between 11:28 and 17:59 UT on 2014 September 10
    in a sit-and-stare mode.
The IRIS slit crossed the main part of a hot channel near its east footpoint.
The formation temperature of the Fe XXI 1354.08 \AA\ line
    is about 11 MK \citep{2013ApJ...763...86L}.
%    \citep{2013ApJ...763...86L, 2015ApJ...799..218Y}
This is the only IRIS line that is formed at a temperature higher than 2 MK
    \citep{2015ApJ...811..139T}.
Here, we use this line to examine the transport of hot plasma
    associated with the hot channel.
%Soft X-ray fluxes were obtained from the GOES,
%    which provides whole-Sun integrated fluxes
%    in two wavebands, 1--8 \AA\ and 0.5--4 \AA.
%\red{This interval has previously been analyzed by \citet{2015ApJ...811..139T},
%    \citet{2015ApJ...813...59L},
%    \citet{2015ApJ...804...82C},
%    and \citet{2016ApJ...823L...4C}.
%\citet{2015ApJ...811..139T} focused on the temporal connection between
%    chromospheric evaporation
%    and the energy deposition rate of the no-thermal electrons during the main phase of the flare.
%\citet{2015ApJ...813...59L} analyzed the chromospheric evaporation
%    and quasi-periodic oscillation of the flare ribbons during the impulsive phase.
%\citet{2015ApJ...804...82C}, and \citet{2016ApJ...823L...4C} investigated
%    the build-up of the magnetic flux rope (MFR).
%}

%For this event,
%    we examine the observations from remote-sensing instruments
%    to identify signatures of relevance to the CME onset.
%We identify plasma heating at the footpoints of the hot channel
%    and transport of hot plasma along a sigmoidal shaped structure
%    in the source region before the flare onset.
To analyze the heating and plasma transportation in more detail,
    we calculate the Differential Emission Measure (DEM)
    using improved tools developed by
    \citet{2015ApJ...807..143C} and \citet{2018ApJ...856L..17S}.
%The improved DEM solutions derived from AIA observations only are more accurate
%    than the previous version,
%    especially for higher temperatures.
%This improvement is vital for the accurate analysis of the flare process \citep{2018ApJ...856L..17S}.
We reconstruct the 3D magnetic field structures of the active region
    using the HMI vector magnetograms using the method of
    \citet{2013ApJ...768..119Z, 2016ApJ...826...51Z}.
    %taken by HMI on SDO.
%Here the magnetohydrostatic model based on the MHD relaxation method is adopted
%    \citep{2013ApJ...768..119Z, 2016ApJ...826...51Z}.

%The fact that there was a CME eruption from the source region
%    being observed by the instruments described above.
%It also was shown by instrumentation aboard the NASA STEREO and ESA/NASA SOHO spacecraft,
%    and the associated in-situ observations relevant to the
%    1 AU arrival of the event is discussed later.
%The instruments include the coronagraphs
%    COR1 (field of view (FOV) of $1.4-4.0 R\_sun$),
%    COR2 (FOV of $2.5-15.0 R\_sun$),
%    and Heliographic Imager 1(HI1, FOV $4.0^{0}-24.0^{0}$ ),
%    Heliographic Imager 2(HI2, FOV $18.7^{0}-88.7^{0}$) onboard STEREO-A and B,
%    and the Large Angle Spectrometric Coronagraphs (LASCO) onboard SOHO.

The speed of the associated CME was very high;
    the linear speed is $\sim$1267 \velunit\ near the Sun
    (based on the CDAW catalog \footnote{https://cdaw.gsfc.nasa.gov/CME\_list/}).
This event is also identified in the HELCATS
    \footnote{https://www.helcats-fp7.eu/index.html}
    catalogue of CMEs observed in Solar TErrestrial RElations Observatory (STEREO)
    Heliographic Imagers data.
The radial speed of the CME derived
    from geometrical fitting to its time-elongation profile
    \citep{2013ApJ...777..167D}
    is $\sim$1000 \velunit.
The CDAW and HELCATS catalogues suggest a transit time to Earth of 2 days.
Consistent with this prediction,
    a CME is detected by the Wind \footnote{https://cdaweb.gsfc.nasa.gov/index.html}
    \citep{1995SSRv...71....5A}
    and Advanced Composition Explorer
    \citep[ACE \footnote{http://www.srl.caltech.edu/ACE},][]{1998SSRv...86....1S}
    spacecraft near the Earth
    during the latter half of September 12.
This is the only CME detected near the Earth within several days of this predicted arrival time.

%To examine the impact of source region activity
%    on the composition parameters of the associated ICME.
%We examine in-situ measurements made by the
%    Wind and ACE spacecraft.
The velocities and number densities of the proton and helium ions of the CME
    were measured
    by the Solar Wind Experiment (SWE) Faraday cup instruments
    \citep{1995SSRv...71...55O}
    onboard Wind. %from which \ahe\ was derived.
{\ahe, expressed as a percentage,
    is derived from the ratio of the number density of helium ions to protons.}
The ICME magnetic field was recorded by the Wind
    Magnetic Field Investigation
    \citep[MFI,][]{1995SSRv...71..207L}.
The average charge state of \qfe\ and the \oql\ ratio were based on measurements
    from the Solar Wind Ion Composition Spectrometer \citep[SWICS,][]{1998SSRv...86..497G}
    onboard ACE.
%Start and end times of ICME and the magnetic cloud (MC) are taken from the HELCATS ICMECAT catalogue.
%This ICME is also included in the WP4 Catalogue \citep{2017SpWea..15..955M}
%    in HELCATS,
%    and the start and end times of the ICME and the magnetic cloud (MC)
%    are taken from the catalogue.

\section{Observational results} \label{sec:results}

\subsection {The morphology evolution and brightness variation before the flare onset}
\label{sec:morphology evolution and brightening variation}
%the subsection can be deleted to make the results more briefly.

\begin{figure*}[ht!]
\begin{center}
%\begin{figure*}[1]    %%%%%%%%%%%%%%%%%% FIGURE 1
%\figurenum{1}
\centerline{\includegraphics[width=0.9\textwidth]
{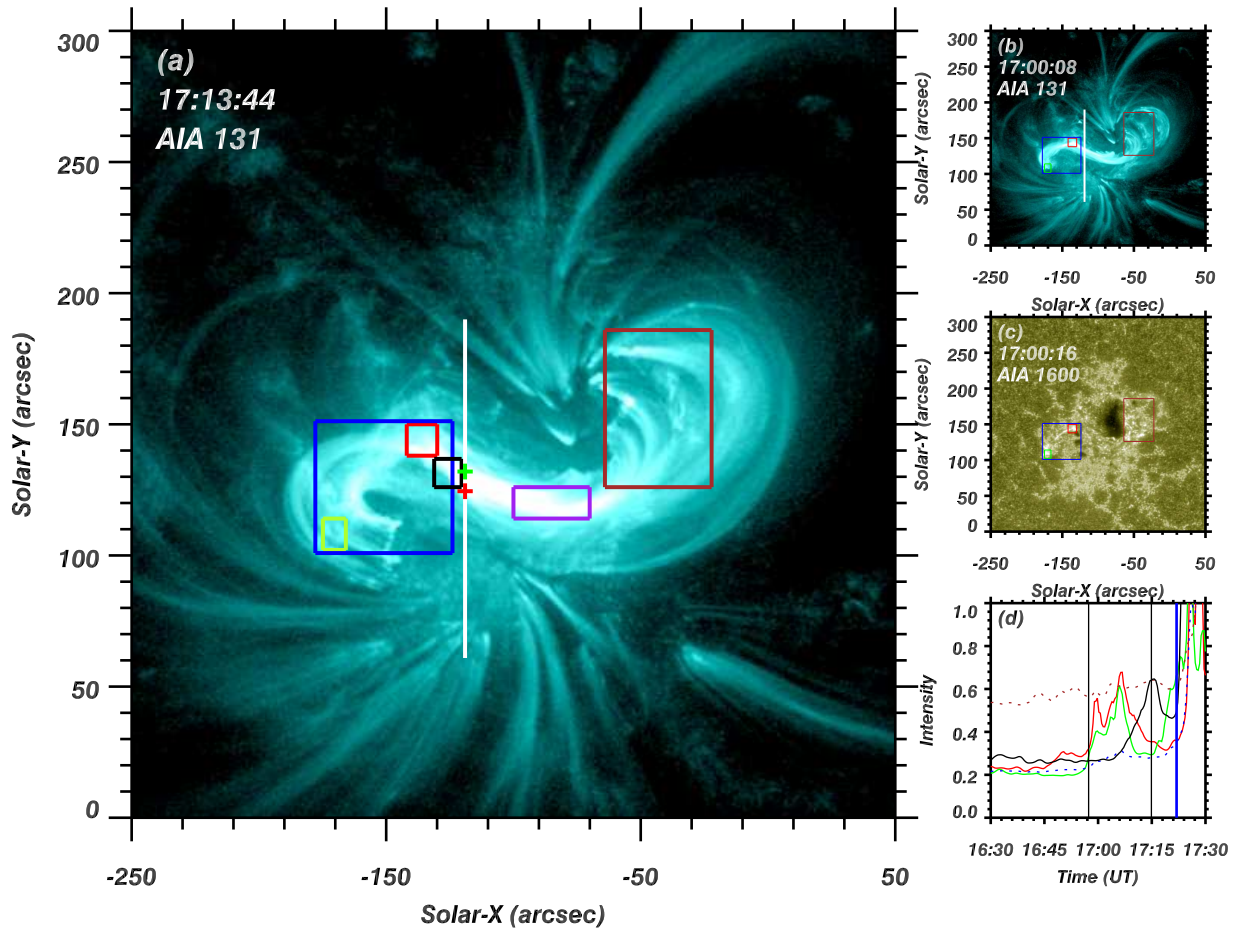}}
\caption{
 Morphology evolution and the footpoint brightenings of the hot channel.
 The evolution of the sigmoidal structure in the higher temperature AIA images (131 \AA)
    is shown in panels (a) and (b),
    and the chromospheric brightening (AIA 1600 \AA) is shown in (c).
 The chromosphere intensity variations of the footpoint regions are shown in (d).
 The white vertical lines in (a) and (b) represent
    the location of the IRIS slit,
    and the green and red plus signs that lie on that line
    denote the north and south loops that are intense in the IRIS Fe XXI spectrum.
 The purple rectangle represents the center part of the hot channel,
    and the other large and small rectangles denote the footpoints of the sigmoidal structure.
% The six coloured rectangles in panel a1 span the sigmoidal structure
%    from the left footpoint to the far right.
% The morphology evolution and the footpoint brightenings of the hot channel.
% The evolution of the sigmoidal structure in the higher temperature AIA images (131 \AA)
%    is shown in a1--a3,
%    and the chromospheric evolution (AIA 1600 \AA) is shown in b1--b3.
% The white vertical lines in panels a1--a3 represent
%    the location of the IRIS slit.
% The large and small rectangles denote the footpoints of the sigmoidal structure.
% The six coloured rectangles in panel a1 span the sigmoidal structure
%    from the left footpoint to the far right.
% The morphology evolution and brightness variation of the source region.
% The evolution of the sigmoidal structure in the higher temperature AIA images (131 \AA)
%    is shown in the top row (a1-a4),
%    and the chromospheric evolution (AIA 1600 \AA) is shown in the middle row (b1-b4).
% The bottom row presents the photosphere magnetic field structure (c1),
%    GOES X-ray intensity (c2),
%    a ST map based on the AIA 193 \AA\ images (c3)
%    (along the location of the slit marked by orange line in (a4)),
%    and the intensity variations in the chromosphere (c4).
% The white vertical lines in panels a1--a4 represent
%    the location of the IRIS slit.
% The large and small rectangles denote the footpoints of the sigmoidal structure,
%    and the chromospheric intensity variations are shown in panel c4 with associated colors.
% The blue vertical lines in panels c2--c4 highlight the time of the flare onset.
 \label{fig:morphology_and_brightening_evolution}}
 \end{center}
\end{figure*}

%An overview of t
Figure~\ref{fig:morphology_and_brightening_evolution}
    shows the morphology evolution and the footpoint brightenings of the hot channel.
The images display a well-developed active region (AR 12158)
    with a large sigmoidal-shaped channel identified in the hot EUV-emitting plasma.
%We concentrate on the morphology and evolution of the sigmoid
%    and the brightening of its footpoint regions prior to the onset of a flare
%    at 17:22 UT.
The sigmoidal structure on AIA 131 \AA\ images existed for $\sim$1
%about an
    hour before CME eruption,
    and is already well-formed by $\sim$17:00 UT (Figure~\ref{fig:morphology_and_brightening_evolution}(b)).
The hot channel (sigmoid) is believed to represent a magnetic flux rope
    \citep[MFR,][]{2012NatCo...3..747Z, 2018ApJ...868...59L},
%    \citep[MFR,][]{2012NatCo...3..747Z, 2013ApJ...763...43C, 2014ApJ...789...93C,
%    2014ApJ...794..149C, 2018ApJ...868...59L},
    which should be mature by this time.
From 17:00 UT,
    the sigmoid structure in AIA 131 \AA\ images (formation temperature $\sim$10 MK)
    becomes increasingly bright, and enlarges.
%Panels a and b show the evolution
%of the morphology
%    of the hot channel morphology
%    in AIA 131 \AA\ images
%    (formation temperature $\sim$10 MK).

%Chromospheric brightening is examined by means of the AIA 1600 \AA\ image in panel (c).
%increases in size.
%To describe the expansion more quantitively,
%    we measure the expansion velocity from AIA 193 \AA\ images
%    by generating a space-time (ST) map
%    (Figure~\ref{fig:morphology_and_brightening_evolution} (c3)).
%The loops that overlie the hot channel are clearly seen in 193 \AA\ images.
%The loops are pushed outward with the expansion of the hot channel
%    (i.e. the sigmoid).
%We can only derive the sideways expansion speed %of the movement
%    from the imaging observations.
%The expansion starts at $\sim$17:00  UT,
%    and the expansion speed is low until flare onset (17:22 UT).
%Thereafter, the expansion speed is much faster.
%This two-phase scenario is the same as the definition
%    for the eruption of flares and CMEs in
%    \citet{2001ApJ...559..452Z}.
During the slow expansion and intensity increase of the sigmoid before the flare onset (17:22 UT),
    chromospheric brightening occurs at the footpoints of the sigmoid
    (brightenings inside blue and brown big rectangles in
    Figure~\ref{fig:morphology_and_brightening_evolution} (c));
    the intensity of those footpoint regions also increase
    in 131 \AA.
The chromosphere intensity variations of the footpoint regions are shown in
    Figure~\ref{fig:morphology_and_brightening_evolution}(d)
    for clarity but will be
    discussed in Section~\ref{sec:IRIS observations}.
These observations demonstrate that the brightening and expansion of the sigmoid
    may be closely related to the brightening of its chromospheric footpoints.
%How are these two phenomena related?
%Do those two phenomena have a causal relationship?
%The above observational results demonstrate that the brightening and expansion of the sigmoid
%    may have a close relationship
%    with the brightening of its footpoint regions in the chromosphere.
%How are these two phenomena related?
%Do they have a causal relationship?

\subsection {The temperature evolution at the footpoints and in the main body of the hot channel}
\label{sec:temperature evolution}

\begin{figure*}[ht!]
\begin{center}
%\begin{figure*}[1]    %%%%%%%%%%%%%%%%%% FIGURE 1
%\figurenum{1}
\centerline{\includegraphics[width=0.9\textwidth]
{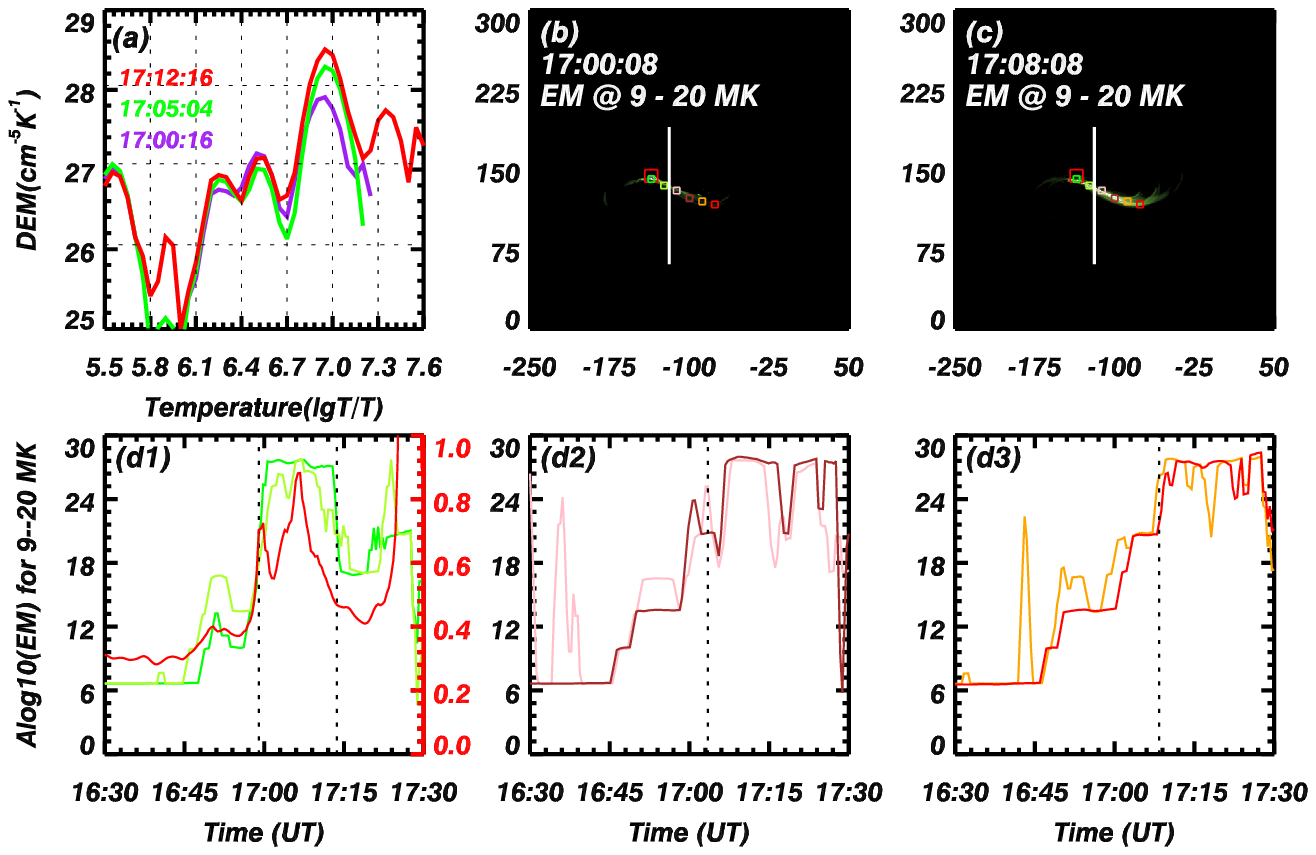}}
\caption{
 The DEM of the hot channel.
 The DEM for the center part of the hot channel is shown in (a).
 The evolution of the EM in the temperature rang from 9--20 MK is shown in (b) and (c).
 The six coloured rectangles span the sigmoidal structure
    from the left footpoint to the far right,
    and the white vertical lines show the location of the IRIS slit.
 The variation of the average EM (9--20 MK) for the six boxes
    is shown in the bottom panels,
    in corresponding colours.
 The chromospheric intensity variations for the footpoint is shown in (d1) with red.
% It is clear that the high-temperature plasma first appears at footpoint
%    and then extends along the sigmoidal structure.
 \label{fig:hot_structure_extending}}
 \end{center}
\end{figure*}

%The Emission Measure (EM) is the volume emission of the plasma.
%The DEM describes the temperature distribution of the plasma
%    as a function of the density of electrons and the gradient of the temperature,
%    and it represents the emission of plasmas along the line of sight
%    at a certain temperature.
%The DEM describes the emission of plasma along the line of sight
%    at a certain temperature.
%We can examine the temperature more accurately by DEM analysis.
%The temperature response functions of AIA measurements are wide
%    \citep{2012SoPh..275...17L}.
%Therefore, the images at each wavelength include radiation
%    from plasma with a wide temperature range.
%IRIS has no spectral observation in coronal temperature
%    except for the Fe XXI 1354 \AA\ line.
%Observations at coronal temperatures from IRIS are limited
%    to the Fe XXI 1354 \AA\ line.
%To examine the temperature evolution of the chromospheric brightenings
%    at the sigmoid footpoint and the extension of the hot structure,
%    we deduce the DEM of the source region.
{The DEM in the center part (purple rectangle in Figure~\ref{fig:morphology_and_brightening_evolution}(a))
    of the hot channel is shown in
    Figure~\ref{fig:hot_structure_extending} (a).
The peak of the distribution is located at $\sim$ 8 MK,
    which demonstrates the high temperature of the hot channel.
During the evolution, the peak of the distribution increased significantly,
    indicating an increase in the amount of hot plasma inside the hot channel.}

%We concentrate on the left footpoint region
%    (small red rectangle
%    in Figure~\ref{fig:morphology_and_brightening_evolution}).
Emission Measure (EM) images with temperature between 9 and 20 MK are shown in
    Figure~\ref{fig:hot_structure_extending}(b) and (c).
High temperature plasma first appears
    at the left (eastern) sigmoid footpoint
    (red rectangle)
    at 16:58:08 UT,
    and then extends along the length of the sigmoid. %for about 40 arcsecs by 17:05:20 UT
%    (i.e. extending at $\sim$70 \velunit).
%The high temperature structure then extends along the sigmoid.
%The farther away from the footpoint,
%    the later the high temperature plasma appears.
%The high temperature structure extends for about 40 arcsecs at 17:05:20 UT.
%Therefore, the deduced speed at which the high temperature structure extends
%    is about 70 \velunit.%(40*725.3 km/390s)
%
%
%In order to
%To evaluate the extension speed of the high temperature structure
%    more accurately,
We choose six rectangles along the sigmoidal structure
    from its left footpoint to the far right
    (see Figure~\ref{fig:hot_structure_extending}(b) and (c)).
The temporal variations of the average EM (9--20 MK) in these six boxes are shown
    in Figure~\ref{fig:hot_structure_extending}((d1)--(d3)),
    in the same colors as the boxes.
The average EM for all six sub-regions is significantly lower than 27 before 16:58 UT
%    (dotted vertical line in Figure~\ref{fig:hot_structure_extending}(d1))
    showing that the plasma temperature % in all six sub-regions
    is lower than 9 MK.
The variation of the EM (9--20 MK), %within the temperature range from 9 MK to 20 MK,
    averaged in the footpoint region
    (green line in Figure~\ref{fig:hot_structure_extending}(d1)) mirrors
    the intensity
    %fluctuation
    increase and decrease in the chromospheric images
    (red line in Figure~\ref{fig:hot_structure_extending}(d1)).
    %add the intensity variation of the footpoint (inside red rectangle)
The chromospheric brightening in the footpoint region and the EM (9--20 MK)
    % from 9 MK to 20 MK
    both start to increase at 16:58 UT before reducing at about 17:14 UT.
The synchronization of these two phenomena
%    EM (9--20 MK) and chromospheric intensity variation
    in this footpoint region means that the plasma was heated to values above 9 MK
    with the brightening in the chromosphere.

Plasma hotter than 9 MK first appears in the leftmost rectangle
    (Figure~\ref{fig:hot_structure_extending}(d1)) at 16:58 UT and in the right-most rectangle,
    50 arcsecs away by 17:08 UT (d3).
%Hot plasma appeared in the other five sub-regions
%    at about 17:01 UT (c1), 17:03 UT (c2), 17:06 UT (c2), 17:07 UT (c3), and 17:08 UT (c3).
%The distance between the most distant sub-regions
%    (the leftmost of which corresponds to the footpoint) is about 50 arcsecs.
Therefore, the hot plasma extends along the sigmoidal structure
    at a speed of $\sim$60 \velunit. %(725*50/600).
%The two methods to deduce the speed at which the hot plasma moves yield almost the same result.
%At this point, the question is:
But does this speed represent the actual flow of the hot material or not?

\subsection {The blueshift of Fe XXI 1354 \AA\ observed by IRIS}
\label{sec:IRIS observations}

\begin{figure*}[ht!]
\begin{center}
%\begin{figure*}[1]    %%%%%%%%%%%%%%%%%% FIGURE 1
%\figurenum{1}
\centerline{\includegraphics[width=0.9\textwidth]
{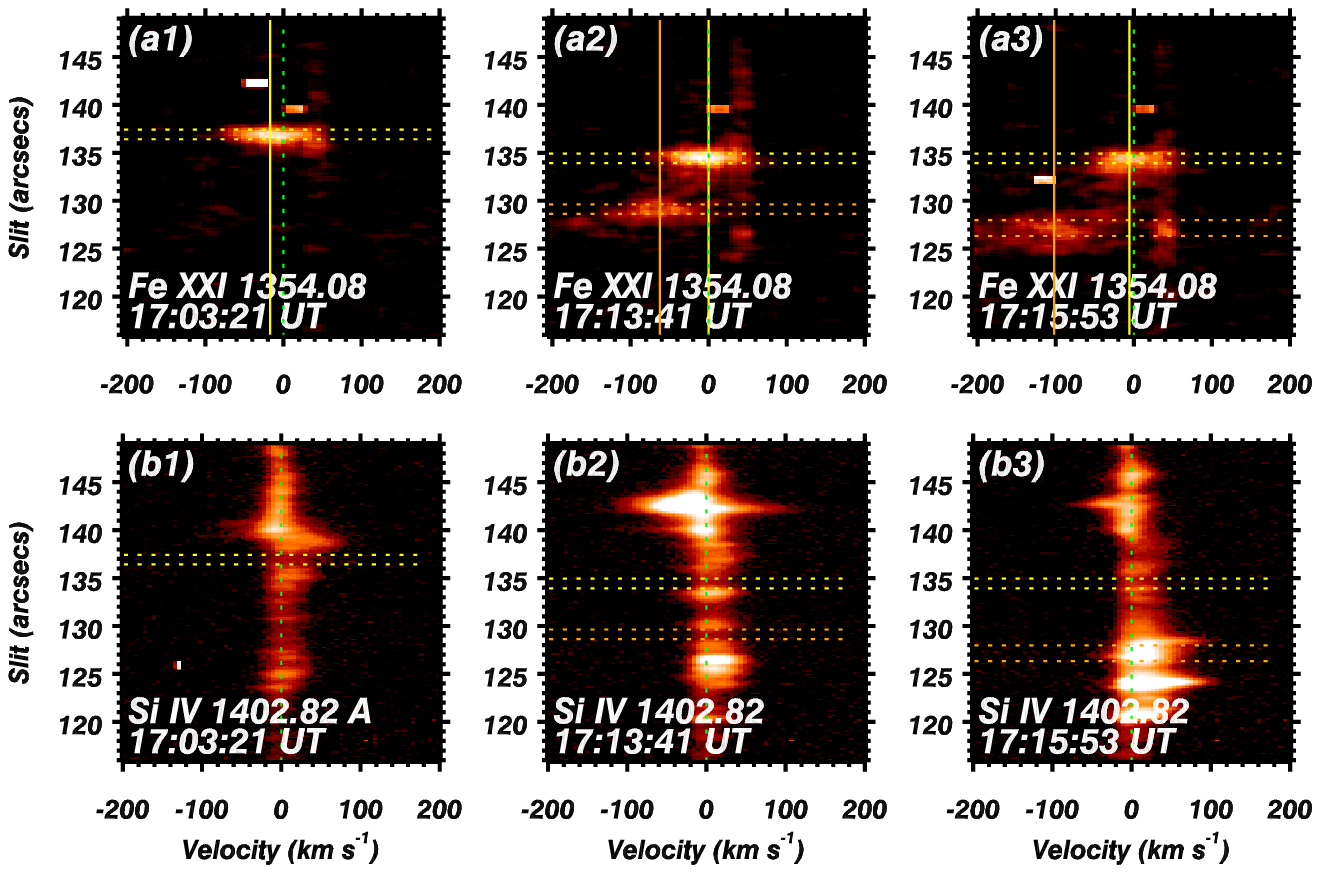}}
\caption{
 The IRIS spectroscopy observations.
 Fe XXI 1354.08 \AA\ and Si IV 1402.82 \AA\ spectra are shown in the top and bottom rows, respectively.
 The green dotted lines represent the location
    where the Doppler shift of the Fe XXI 1354.08 \AA\ line is zero.
 The intensity of the regions between the pairs of yellow and golden horizontal lines
    in the Fe XXI 1354.08 \AA\ spectrum
    is higher than elsewhere,
    and the line centers were deduced from measurements in those regions.
 The yellow and golden solid vertical lines in the top panels
    denote the deduced line centers for Fe XXI 1354.08 \AA\
    for northern and southern loop systems, respectively.
 \label{fig:IRIS_observations}}
 \end{center}
\end{figure*}

%Fortunately, we have the spectroscopy observations taken by the IRIS for this interval.
The IRIS slit
% which is represented by the
    (white vertical line in
    Figure~\ref{fig:morphology_and_brightening_evolution}
    and Figure~\ref{fig:hot_structure_extending}),
    was over the main part of the sigmoid
    near its eastern footpoint.
%The intensity of the Fe XXI 1354.08 \AA\ emission starts to increase at 17:01 UT,
%    which is almost the same
%    close to the time at which the EM in the green yellow rectangle in Figure~\ref{fig:hot_structure_extending}(c1) starts to increase.
%As the IRIS slit is located near this green rectangle
%    (Figure \ref{fig:hot_structure_extending}),
%    this means that the higher temperature structure deduced by the DEM method is credible.
%
%
%The
IRIS Fe XXI 1354.08 \AA\ spectra taken at different times
    are shown in Figure~\ref{fig:IRIS_observations}(a1)--(a3);
    Figure~\ref{fig:IRIS_observations}(b1)--(b3) present
    spectra from Si IV 1402.82 \AA.
The C I 1354.288 \AA\ lies close to the Fe XXI 1354.08 \AA\ in the spectrum,
%This C I line is much narrower than the Fe XXI line
    but is much narrower as it emitted from neutral atoms
    and its Doppler shift %corresponds to a small redshift, or nearly zero
    is nearly zero
    \citep{2015ApJ...811..139T}.
Therefore, we use this line %as the reference by which
    to calibrate the Fe XXI 1354.08 \AA.
%The green vertical dotted lines in Figure~\ref{fig:IRIS_observations}
%    represent the location where the Doppler shift of Fe XXI 1354.08 \AA\ is zero.
%There are two places with high intensity in the Fe XXI 1354.08 \AA\ spectra,

{The Fe XXI 1354.08 \AA\ spectrum is intense near solar y=135 and 125 arcsecs.
These positions are delimited by pairs of horizontal yellow and golden lines in Figure~\ref{fig:IRIS_observations}, respectively, and marked by the two plus signs in
    Figure~\ref{fig:morphology_and_brightening_evolution}(a).}
%The Fe XXI 1354.08 \AA\ profiles within each pair of horizontal lines
%    are averaged to produce the profile peak,
%    which are plotted as vertical lines of the corresponding color.
%The deduced Doppler shifts of these peaks at 17:03 UT and 17:13 UT
%    is almost the same as was found by \citet{2016ApJ...823L...4C}.
%There are two characteristic signatures of the Fe XXI 1354.08 \AA\ spectra.
{The Fe XXI line intensity starts to increase at 17:01 UT at 135 arcsecs; %
    %in the north part of the slit
%    (the footpoint locates at the small red rectangel in
%    Figure~\ref{fig:morphology_and_brightening_evolution}(a));
    later, from about 17:11 UT,
    the intensity in the south part of the slit
%    (the footpoint locates at the small black rectangel in
%    Figure~\ref{fig:morphology_and_brightening_evolution}(a))
    starts to increase.}
In both locations, the Fe XXI line is blueshifted, and
%The Doppler speed of the north loop is about 10--20 \velunit\
%    at 17:03 UT (yellow vertical line in Figure~\ref{fig:IRIS_observations}(a1)),
%    reducing to $\sim$0 after 17:10 UT
%    (yellow vertical lines in Figure~\ref{fig:IRIS_observations}(a2) and (a3)).
%In contrast, the blueshift is much higher for the south loops,
%    reaching speeds up to 100 \velunit\
%    (golden vertical lines in Figure~\ref{fig:IRIS_observations}(a2) and (a3)).
    the Doppler speed of the north (south) loop is about 10--20 (100) \velunit.
This demonstrates that higher temperature plasma
    propagates from lower to higher altitudes in the north and south loops systems.

We believe that the hot plasma flows in both
    north and south loops come from the footpoint regions
    represented by red and black boxes in
    Figure~\ref{fig:morphology_and_brightening_evolution}(a).
%The reasons are as the follows.
First, the high temperature plasma first appears at the sigmoid footpoint
    and then extends along the sigmoidal structure
    (Figure~\ref{fig:hot_structure_extending}(b) and (c)).
%    to the north loop system (see Figure~\ref{fig:hot_structure_extending}(b)--(c)).
%The time range over which the Fe XXI 1354.08 \AA\ line is blueshifted
%    in the north loop system (17:01--17:10 UT)
%    coincides with the brightening due to plasma heating in the red box
%    (16:58--17:10 UT, see red line in Figure~\ref{fig:morphology_and_brightening_evolution}(c4)).
%The 1354 \AA\ line becomes blueshifted several minutes after
%    the footpoints brighten.
%This lag should correspond to the propagation time of the hot plasma
%    along the sigmoidal structure from the footpoint to the IRIS slit.
Second, the time ranges over which the Fe XXI 1354.08 \AA\ line is blueshifted
     in the north (16:58--17:13 UT) and south (17:12--17:18 UT) loop systems coincide with
     the brightening due to plasma heating in the red and black boxes (Figure~\ref{fig:morphology_and_brightening_evolution}(d)), respectively.
%the footpoint of the southern loop system
%    (black rectangle in Figure~\ref{fig:morphology_and_brightening_evolution} (a) and (c))
%    starts to brighten
%    in the chromosphere at around 17:11 UT
%    (Figure~\ref{fig:morphology_and_brightening_evolution}(d)).
%The 1354.08 \AA\ line in the southern loop also begins to brighten at $\sim$17:11 UT,
%    with a strong blueshift which is evident until 17:18 UT.
%This time range %is almost the same as
%    almost matches that of the chromospheric brightening
%    of the southern loop system footpoint region (17:11--17:18 UT),
%    the intensity variation of which is shown by the black line
%    in Figure~\ref{fig:morphology_and_brightening_evolution}(c4).

Furthermore,
    we examine the chromospheric lines at the location where the Fe XXI 1354.08 \AA\ line
    is intense and blueshifted.
The results confirm those of \citet{2016ApJ...823L...4C},
    in that the chromospheric lines do not brighten, broaden, or shift between 17:01 UT and 17:14 UT
    (see Figure~\ref{fig:IRIS_observations} (b1) and (b2)), %(17:14:30)
    meaning that the hot plasma up-flow %observed in Fe XXI 1354.08 \AA\
    is not generated
    in the low atmosphere where the IRIS slit is located.

%The question to be asked is
%So why are the Doppler shift of the northern and southern loop systems
%    significantly different
%    ($\sim$10 compared with $\sim$100 \velunit)?
%The hot plasma is transported along the sigmoid from the footpoints
%    for both the northern and southern loop systems.
%The speed deduced from AIA images corresponds to the
%    propagation of the hot plasma projected onto the sky plane,
%%In contrast,
%    whereas the Doppler shift represents the line-of-sight (LOS) speed.
%For the northern loop system,
%    the distance between the IRIS slit and the associated footpoint
%    is farther than that for the southern loop system,
%    (see Figure~\ref{fig:morphology_and_brightening_evolution}(a4)).
%%The slit is close to the brightening footpoint regions for the south loop system.
%Therefore, the northern loop is directed close to the sky plane
%    at the location of the slit,
%    as the angle between the magnetic field lines of the sigmoid
%    and the sky plane decrease with %increase in altitude.
%    increasing altitude.
%In contrast,
%    the direction of the hot plasma flow should be close to the LOS
%    for the southern loop system as the IRIS slit is near the associated footpoint.
%If the speed of the hot plasma flows along the sigmoid %are both
%    was hundreds of \velunit,
%%    in the northern and southern loop systems,
%    the observed Doppler shift of the southern loop system would be much higher.
%%    than that for the northern loop system.

Therefore, the spectroscopic observations confirm that the higher temperature plasma
    %(hotter than 9 MK)
    generated at sigmoid footpoints
    can propagate along with the sigmoid structure
    (with $\sim$100 \velunit\ speed) and,
    moreover, supply mass to the CME
    due to the fact that the sigmoid (hot channel) represents the helical MFR
    \citep{2012NatCo...3..747Z, 2018ApJ...868...59L}.
%    \citep{2012NatCo...3..747Z, 2014ApJ...789L..35C, 2014ApJ...794..149C, 2018ApJ...868...59L}.
{
%How the hot plasma generated at the footpoint of the sigmoidal structure?
%Magnetic field reconnection or other processes?
}

We believe that the plasma at the sigmoid footpoints
    is heated by chromospheric evaporation,
    not by magnetic reconnection at the footpoint directly,
    for the following reasons:

%First,
%    brightening in the different footpoints,
%    as a whole (blue and brown dotted lines in %Figure~\ref{fig:morphology_and_brightening_evolution}(d)),
%    and in different sub-regions of the same footpoint
%    (green and red small rectangles in Figure~\ref{fig:morphology_and_brightening_evolution}(a)
%    and lines in
%    Figure~\ref{fig:morphology_and_brightening_evolution}(d)) occur near-simultaneously.
%Such brightening consistency in different areas
First, the average intensities of the right and left footpoint regions
    (blue and brown dotted rectangles in Figure~\ref{fig:morphology_and_brightening_evolution}(a)
    and dotted lines in Figure~\ref{fig:morphology_and_brightening_evolution}(d))
    are both increasing gradually during the slow expansion of the sigmoid.
The brightening in different sub-regions of the same footpoint
    (green and red small rectangles in Figure~\ref{fig:morphology_and_brightening_evolution}(a)
    and lines in Figure~\ref{fig:morphology_and_brightening_evolution}(d)) even
    occur near-simultaneously.
This is hard to explain by magnetic reconnection.
%The evolution of brightening in different footpoint areas have good time consistency.
%The time consistency is not only valid for the east and west footpoint regions
%    as a whole (blue and brown dotted lines in %Figure~\ref{fig:morphology_and_brightening_evolution}(c4)).
%Even the intensity variations of the two separated footpoint sub-regions
%    (the green and red small rectangles in %Figure~\ref{fig:morphology_and_brightening_evolution}(b1-b4))
%    in the east footpoint region are also almost the same
%    (green and red lines in Figure~\ref{fig:morphology_and_brightening_evolution}(c4)).
%The so good brightening time consistency for different areas
%    is hard to explain by the magnetic reconnection.
Second, direct evidence for chromospheric evaporation
    is provided by spectroscopic observations taken after 17:14:30 UT.
When the area of chromospheric brightening
    extends to the IRIS slit,
    the intensities of the Fe XXI 1354.08 \AA\ and Si IV 1402.82 \AA\ lines are increased
    in the southern loop system.
The former is strongly blueshifted,
    and the later,brightened, broadened, and redshifted
    (Figure~\ref{fig:IRIS_observations}(a3) and (b3)).
The scenario is the same as for the standard chromospheric evaporation process,
    except that the rate of evaporation is weaker than for the impulsive phase of the flare.
%The blueshift in the southern loop system is higher than it is several minutes previously
%    (see golden vertical lines in Figure~\ref{fig:IRIS_observations}(a2) and (a3)),
%    as the IRIS slit is located in the region
%    where chromospheric evaporation is occurring.
%If the explanation of chromosphere evaporation is valid,
%    the question is where and how the higher energy non-thermal electrons are generated?

\subsection {The overall magnetic field structure of the active region}
\label{sec:magnetic field structure}

\begin{figure*}[ht!]
\begin{center}
%\begin{figure*}[1]    %%%%%%%%%%%%%%%%%% FIGURE 1
%\figurenum{1}
\centerline{\includegraphics[width=0.9\textwidth]
{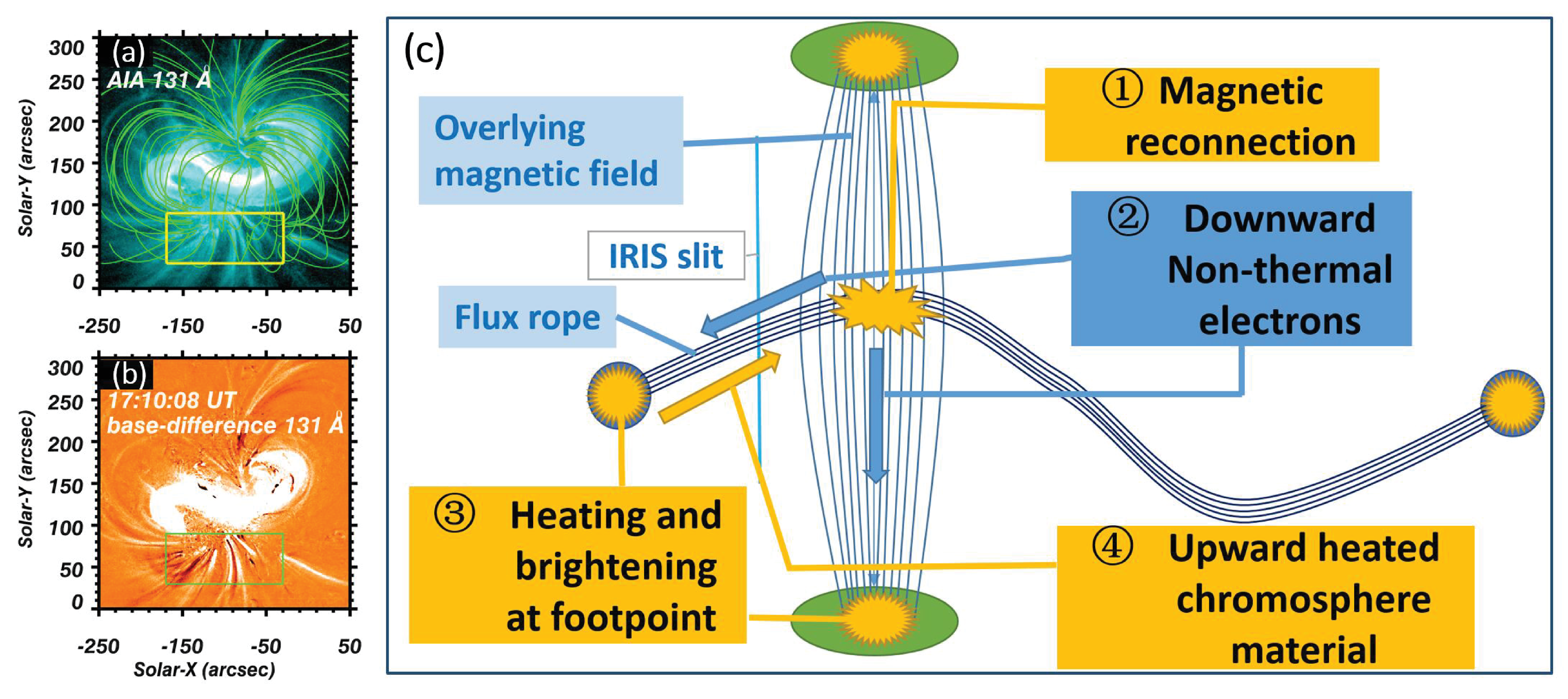}}
\caption{
 The overall magnetic field structure of the active region is shown in (a).
 Base-difference 131 \AA\ image
    is shown in (b).
 Rectangles enclose the footpoint region of the overlying magnetic field structure.
 The scenario for magnetic reconnection between the rising flux rope
    and the overlying magnetic field structure,
    downward non-thermal electrons,
    chromospheric evaporation at the footpoint,
    and up-flowing heated plasma are shown in (c).
% The intensity of the overlying magnetic field structure footpoint region
%    also increases prior to flare onset.
 \label{fig:magnetic_field_and_cartoon}}
 \end{center}
\end{figure*}

To explore where and how the higher energy non-thermal electrons are generated,
    we show the magnetic field structures
    of the active region in Figure~\ref{fig:magnetic_field_and_cartoon}(a).
    %deduced by magnetohydrostatic modeling
    %through the MHD relaxation method
    %\citep{2013ApJ...768..119Z, 2016ApJ...826...51Z},
    %based on the vector magnetograms
    %taken by HMI.
%\red{The magnetic field structure of the sigmoid is described in more detail by
%    \citet{2018ApJ...868...59L}.}
%It is clear that
The sigmoid is clearly covered by overlying magnetic field structures.
The %picture illustrates a
    scenario of the whole process
    is illustrated in Figure~\ref{fig:magnetic_field_and_cartoon}(c).
{High-energy, non-thermal electrons, produced by magnetic reconnection between
    the rising flux rope and the overlying magnetic field structure (step 1 in
    Figure~\ref{fig:magnetic_field_and_cartoon}(c)),
    are transported downwards to the sigmoid footpoints (step 2).
These non-thermal electrons then cause heating of the chromospheric material (step 3).
The heated chromospheric material is then transported upwards into the sigmoidal structure
    along the magnetic field lines (step 4).}
%    which is the same as that for {\bf magnetic} reconnection
%    between the rising flux rope and overlying field lines
%    \citep[e.g.,][]{2019ApJ...878...38Y}.
%This %picture illustrates a
%    scenario is illustrated in Figure~\ref{fig:magnetic_field_and_cartoon}(c).
%    \citep[e.g.,][]{2001ApJ...552..833M, 2001JGR...10625227S, 2004ApJ...602.1024S, %2004ApJ...611..545G, 2019ApJ...878...38Y}.
%
%A cartoon showing the primary processes associated with the magnetic reconnection
%    between the rising flux rope and the overlying magnetic field structure
%    is shown in Figure~\ref{fig:magnetic_field_and_cartoon}(a3).
%Such a scenario is illustrated in Figure~\ref{fig:magnetic_field_and_cartoon}(a3).
The observed phenomena,
    such as the time correspondence of footpoint brightening,
    heating in the footpoint regions,
    and high temperature plasma up-flow,
    along with the presence of the sigmoidal structure,
    are all consistent with the picture.
    %scenario as shown in Figure~\ref{fig:magnetic_field_and_cartoon}(a3).

In this scenario,
    the intensity of the footpoints of the overlying magnetic field structure
    should also increase \citep{2019ApJ...878...38Y}. % due to the
%    high energy non-thermal electrons produced by magnetic reconnection
%    between the rising flux rope and the overlying magnetic field
%    \citep{2019ApJ...878...38Y}.
A base-difference 131 \AA\ image is presented in
    Figure~\ref{fig:magnetic_field_and_cartoon}(b).
Note that the intensity of the overlying magnetic field structure %southern
    footpoint region
    (green rectangles)
    was also enhanced. % between 16:50 and 17:20 UT (prior to flare onset).
This supports the scenario illustrated
    in Figure~\ref{fig:magnetic_field_and_cartoon}(c).
%our suggestion
%    that the hot plasma inside the CME is heated by chromospheric evaporation
%    at the sigmoid footpoints.

\subsection {The in situ properties of the ICME} \label{sec:results in situ properties}

\begin{figure*}[ht!]
\begin{center}
%\begin{figure*}[1]    %%%%%%%%%%%%%%%%%% FIGURE 1
%\figurenum{1}
\centerline{\includegraphics[width=0.9\textwidth]
{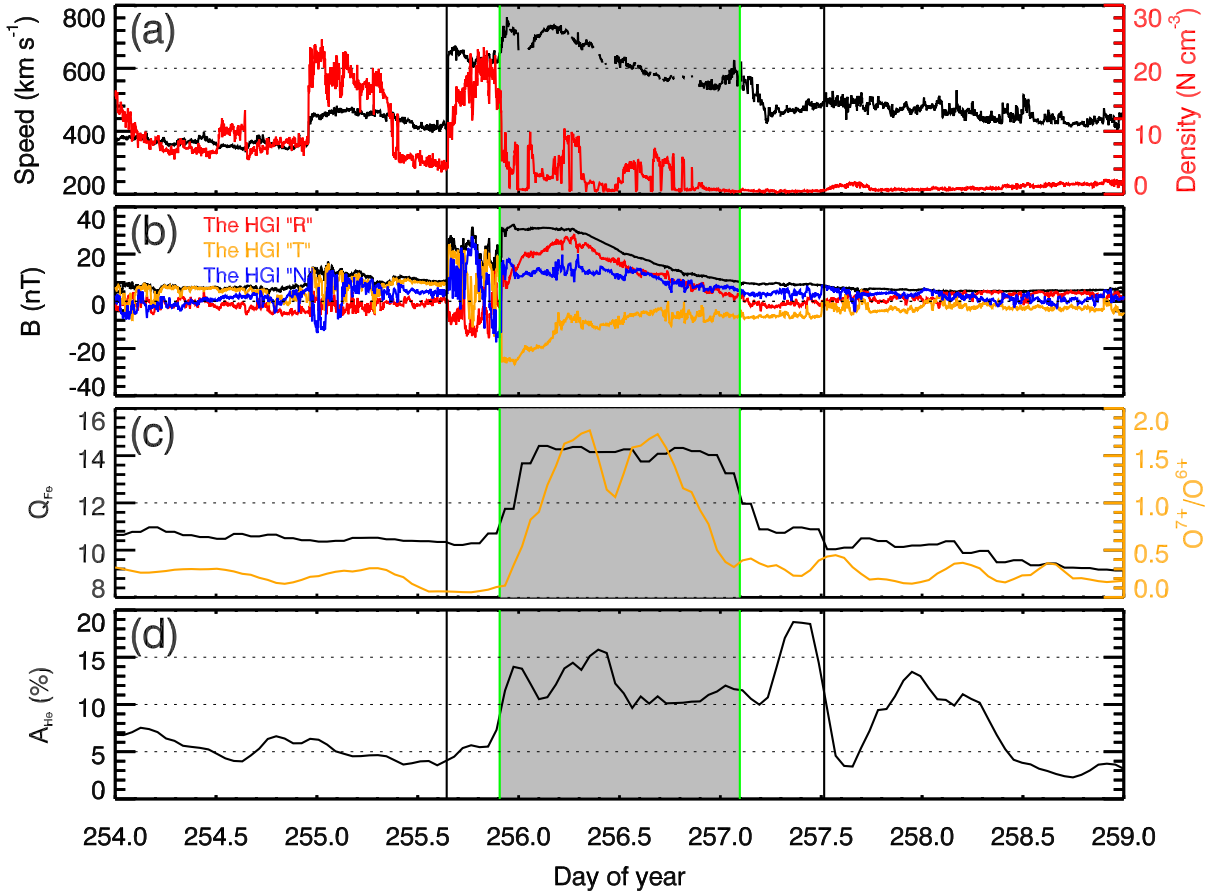}}
\caption{
 The in-situ parameters of the ICME and the background solar wind.
 The speed and density are shown in (a),
    magnetic field in (b),
    average charge state of Fe and density ratio of $O^{7+}$ and $O^{6+}$ in (c),
    and helium abundance in (d).
 The two black vertical lines represent the start and end of the ICME passage and
    the MC is represented by the grey region bounded by green vertical lines.
 The charge state of \qfe\ and the \oql\ are much higher inside the MC than elsewhere
    in the ICME/in the solar wind,
    and the \ahe\ is also as high as it is in the photosphere.
 \label{fig:in_situ_parematers}}
 \end{center}
\end{figure*}

Figure~\ref{fig:in_situ_parematers} shows the in situ parameters of the associated ICME
    from Wind and ACE.
%The two black vertical lines represent the start and end of the ICME,
%    and the edges of the MC are represented by the two green vertical lines.
%The magnetic field is typical for a MC.
%The ICME took around two days to pass over the spacecraft,
%    and the MC lasted for $\sim$29 hours.
There is a high density sheath in front of the magnetic cloud (MC),
    because the ICME speed (600 \velunit)
    was much higher than the background solar wind (400 \velunit).
%It is worth pointing out
Note that the charge states and \ahe\
    in the sheath are not elevated;
    they are almost the same as in the upstream solar wind,
    indicating that the sheath is composed of swept-up material.
The average \qfe\ (\oql) reaches 15 (1.5) inside the MC,
    but is only $\sim$10 (0.2)
    in the surrounding background solar wind,
    indicating that
    the source temperature of the ICME is much higher
    than that of the background solar wind.
%An interesting result is that \ahe\ is extremely high.
Interestingly,
    \ahe\ inside the MC is $\sim$10 to 15\%,
    which is even higher than the average \ahe\ in the photosphere.
%Where does the extremely hot and high \ahe\ material
%    inside the MC come from?
%The question is closely connected with the activity within the source regions.

%\red{
%Why is the helium abundance so high inside MC?
%Where does the extremely hot and high helium abundance material
%    inside the MC come from?
%How are the elements inside the MC heated?
%The questions are closely connected with the activity within the source regions.
%}

\section{Discussion}
\label{sec:discussion}

\subsection{Why is the helium abundance so high in ICMEs?}
\label{sec:discussion helium abundance}

It is not known why \ahe\ is so high in some ICMEs
    \citep{1997GMS....99..245N, 2017SSRv..212.1159M}.
\citet{1997GMS....99..245N} speculated that a "sludge removal phenomenon"
    transports lower atmosphere material with higher \ahe\ into the heliosphere
    together with the CMEs.
%In the present study,
Here,
    we find that chromosphere material heated by chromospheric evaporation
    at the footpoints can be transported into the sigmoidal structure
    before flare onset.
It is believed that the sigmoid represents the helical MFR
    \citep{2012NatCo...3..747Z, 2018ApJ...868...59L}
%    \citep{2012NatCo...3..747Z, 2013ApJ...763...43C, 2014ApJ...789L..35C, 2014ApJ...794..149C, %2018ApJ...868...59L}
    and corresponds to the MC inside ICME.
Therefore, in our case,
    the hot plasma inside the ICME comes from the chromosphere
%\red{    and is heated by chromospheric evaporation}
    before flare onset.
The time scale for the FIP bias effect
    is several days
    \citep{2013ApJ...778...69B},
    %hours to days
%    \citep{2001ApJ...555..426W, 2013ApJ...778...69B}.
    whereas chromospheric evaporation is much faster.
{If He (ionization temperature $\sim$50,000K) is still neutral in the chromosphere
    prior to being heated,
    the \ahe\ of the heated plasma should also be high,
    equal even, to its value in the photosphere.}
The aforementioned scenario for the mass supply of the CME
%    can \red{give a reasonable explanation for}
    can explain the high \ahe\ and charge states inside ICMEs.

%\subsection{Improving understanding of chromospheric evaporation}
%\label{sec:improve the understanding of the chromospheric evaporation process}

%We know little about the mechanisms that cause the FIP bias effect.
%Generally,
%    it is believed that some processes are more effective at transporting ions
%    to the upper atmosphere.
%This means that the element abundance should be
%    equal to that in the photosphere
%    if the element is still neutral.
%In the present study,
%Here,
%    we find that \ahe\ is much higher, and the FIP bias is also higher,
%    inside the MC.
%This means that He is still neutral and the low FIP elements,
%    such as Mg, Si, and Fe,
%    should have lost electrons before being heated. %the heating.
%The results indicate that the original temperature of the material
%    heated by chromospheric evaporation process is higher than
%    the ionization temperature of Fe ($\sim$10,000K)
%    and lower than that of He ($\sim$50,000K).
%The results should be useful for theoretical and modeling
%    work on chromospheric evaporation.

%\subsection{The connection between the evolution pre- and post-flare onset
%    and the relationship between the flares and CMEs}
\subsection{The connection between the evolution pre- and post-flare onset}

\label{sec:connection between before and after the flare
    and the relationship between the flares and CMEs}

%In the present study,
%Here, we concentrate on the variation of footpoint brightening and morphology evolution
%    of the hot channel before flare onset,
%The evolution in this phase
    %which
%The present study is not only crucial
%    for understanding the supply of mass to CMEs,
%    but also for analyzing the trigger mechanisms of flares and CMEs
%    and the relationship between these two phenomena.
%The period before flare onset corresponds to
%    the precursor phase of flares and the initiation phase of CMEs.
%\citet{2001ApJ...559..452Z} %analyzed the connection between the kinematic evolution of CMEs
    %and variations in X-ray intensity.
%    found that the initiation phase of CMEs occurred several tens of minutes earlier than
%    the onset of flares.
%    found that flare onset heralds the end of the CME initiation phase % of a CME
%    and the start of the CME impulsive phase;
%    the kinematic parameters of a CME are very different in these two phases.
%Therefore,
%    \citet{2001ApJ...559..452Z}
%    argued that the initiation and impulsive phases represent
%    different phases in CME evolution.
%During the initiation phase,
%    the magnetic field structure undergoes quasi-steady evolution,
%    until it reaches the critical point of flare onset
%    after which the magnetic field structures evolves much faster and more violently.
\citet{2001ApJ...559..452Z} speculated that the development of the physical process
    that occurs before the flare onset is what
    results in the eruption of CMEs and flares.
%\citet{2001ApJ...552..833M} speculated that internal reconnection is critical
%    to the onset of flares, and to eruption of flares and CMEs,
%    as it begins during early on.
%Because of limitations in the observations at that time,
%    it is impossible to analyze the initiation phase in more detail.
%\red{Conversely, \citet{2018ApJ...859..148W} analyzed the very early stages of hot channels,
%    before flare eruption.
%They identified movement of heated plasma in the precursor phases,
%    at speeds of around 1--2 hundred \velunit.
%They suggested that the hot channels result from magnetic reconnections
%    during the precursor stages.
%However, they did not provide more details on how the hot material was heated
%    and where the material came from.}
Here,
    we present a more detailed picture of the connection between flares and CMEs
    in the precursor phase
    (as shown in Figure~\ref{fig:magnetic_field_and_cartoon}(c)).
Magnetic reconnection
    between the flux rope and the overlying loops
    is triggered by the rising of the flux rope.
%    which may be caused by the rotation of the sunspot in the flux-rope footpoint regions.
%High-energy, non-thermal electrons can be produced by the magnetic reconnection,
%    and transported down along magnetic field lines to the sigmoid footpoints.
%These non-thermal electrons then cause heating of the chromospheric material,
%    which is then transported upwards into the sigmoidal structure
%    along the magnetic field lines.
%The sigmoid brightens and further expands.
%This expansion increases the rate of magnetic reconnection.
Non-thermal electrons produced by the magnetic reconnection
    cause heating of the chromospheric material,
    which is then transported upwards into the sigmoidal structure.
The sigmoid brightens and further expands.
This expansion increases the rate of magnetic reconnection.
Chromospheric evaporation that occurs at the footpoint and the magnetic field reconnection
    that occurs at higher altitudes exacerbate each other.
%In the present case, this scenario should be valid before the flare onset at 17:22 UT.
This system evolves much faster after the onset of the flare,
    and the standard CME and flare model should be valid during the impulsive phase.
Our findings demonstrate that evolution before flare onset
    is crucial for the eruption of the flare and CME.
Flares and CMEs have a close relationship,
    and they are different manifestations %(both consequences) of
    of the same active magnetic field system
    \citep{1995A&A...304..585H, 2003AdSpR..32.2425H}.

%Chromospheric evaporation may be responsible for the soft X-ray variation
%    before flare onset,
%    as the moderate brightening at the sigmoid footpoints
%    (blue and brown dotted lines in Figure~\ref{fig:morphology_and_brightening_evolution}(c4))
%    is almost the same as the variation of GOES X-ray flux
%    (Figure~\ref{fig:morphology_and_brightening_evolution}(c2)).
%    and the internal reconnection dominated at this phase.
%The observations demonstrate that
%    flares and CMEs are intimately related;
%    they are different manifestations (both consequences)
%    of the same active magnetic field system
%    \citep{1995A&A...304..585H, 2003AdSpR..32.2425H}.

\section{Conclusion} \label{sec:conclusion}

We have examined the relationship between the in-situ parameters of the ICME
    and activity on the Sun.
In the source region, we focus on
    the connection between sigmoid evolution
    and chromospheric brightening at the sigmoid footpoints
    before flare onset. We find that:
%The main results are summarized as follows:

\begin{enumerate}
\item
During the slow expansion and intensity increase of the sigmoid,
    the high temperature plasma first appears at the footpoints of the sigmoid,
    associated with chromospheric brightening.
Then the high-temperature region extends along the sigmoid.

\item
The spectroscopic observations demonstrate that
    the extension of the higher temperature structure is caused by the upflowing hot plasma
    and the material at the sigmoid footpoints
    are heated by chromospheric evaporation.

\end{enumerate}
%上面部分也可以不加item，直接写下来，但是觉得太不成体系了。有点乱。

%\begin{enumerate}
%\item
%We find that,
%    during the slow expansion and intensity increase of the sigmoid,
%    the sigmoid footpoints also brighten.
%With the help of DEM analysis,
%    we find that higher temperatures first appeared in the sigmoid footpoints,
%    and then extended along the sigmoid structure.

%\item
%The Fe XXI 1354.08 \AA\ line measured by IRIS shows blueshifts
%    ranging from several tens \velunit\ to $\sim100$ \velunit,
%    which demonstrates that the extension of the higher temperature structure
%    is caused by the upflowing hot plasma.
%
%\item
%The spectroscopic observations also show that the Fe XXI 1354.08 \AA\ line
%    is strongly blueshifted, whereas the Si IV 1402.82 \AA\ line
%    brightens, broadens, and is redshifted
%    when the slit is located over the area of chromospheric brightening.
%The IRIS spectroscopic observations and remote-sensing data from AIA and HMI
%    demonstrate that the material at the sigmoid footpoint regions
%    are heated by chromospheric evaporation.
%
%\item
%The footpoints of the overlying magnetic field structures also brighten,
%    which provides additional evidence that
%    the higher energy non-thermal electrons are produced by magnetic reconnection
%    between the rising flux rope and the overlying magnetic field lines.
%
%\end{enumerate}
%上面部分也可以不加item，直接写下来，但是觉得太不成体系了。有点乱。

Our results demonstrate that chromosphere material
    can be heated to higher than 9 MK
    by chromospheric evaporation
    at the sigmoid footpoints before flare onset.
The heated chromospheric material can be transported into the sigmoidal structure
    and supply mass to the CME.
This mass supply scenario provides
    a reasonable explanation for the high charge states and elevated \ahe\
    of the associated ICME.

\acknowledgements{
%{  The authors thank very much the anonymous referee for the helpful comments and suggestions.}
%The authors thank the anonymous referee for helpful comments and suggestions.
%The authors gratefully acknowledge the valuable comments and suggestions made by Dr. R.T. %Wilson, Prof. D. Berkvens, Dr. J. Dixon and Prof. D.U. Pfeiffer on earlier drafts of this %document.
%We acknowledge Drs. Bernhard Kliem and Rui Liu for
%valuable suggestions and comments to improve the study. We
%are grateful to Dr. Yang Guo for providing the code to perform
%the topological study and helpful discussions.
The authors thank the %anonymous
referee for helpful comments and suggestions.
We thank Prof. Jie Zhang for % valuable
    his useful discussions. %comments and suggestions to improve the study.
We thank the ACE SWICS,
SWEPAM, and MAG instrument teams and the ACE Science
Center for providing the ACE data.
Analysis of Wind SWE observations is supported by NASA
    grant NNX09AU35G.
SDO is a mission of NASA’s Living With a Star Program.
IRIS is a NASA small explorer mission developed and
operated by LMSAL with mission operations executed at
NASA Ames Research Center and major contributions to
downlink communications funded by ESA and the Norwegian
Space Centre.
%{STEREO/HI was developed by a
%consortium comprising Rutherford Appleton Laboratory, and
%University of Birmingham (UK), Centre Spatiale de Li`ege
%(Belgium), and the Naval Research Laboratory (USA).}
%The Heliospheric Imager (HI) instruments were developed by a
%collaboration that included the Rutherford Appleton Laboratory
%and the University of Birmingham, both in the United Kingdom,
%and the Centre Spatial de Liège (CSL), Belgium, and the US
%Naval Research Laboratory (NRL), Washington DC, USA. The
%STEREO/SECCHI project is an international consortium of the
%Naval Research Laboratory (USA), Lockheed Martin Solar and
%Astrophysics Laboratory (USA), NASA Goddard Space Flight
%Center (USA), Rutherford Appleton Laboratory (UK), University
%of Birmingham (UK), Max-Planck-Institut für Sonnensystemforschung
%(Germany), Centre Spatial de Liège (Belgium),
%Institut d’Optique Théorique et Appliquée (France), and Institut
%d’Astrophysique Spatiale (France).
This research is supported by
    the National Natural Science Foundation of China (U1931105, 41974201, 41627806, 41604147).
}

\bibliographystyle{aasjournal}
%\bibliography{reference_1,reference_2}
\bibliography{Flares_and_CMEs,Kinematics_and_Morphology_of_CMEs,source_of_CMEs}

\begin{thebibliography}{}
\expandafter\ifx\csname natexlab\endcsname\relax\def\natexlab#1{#1}\fi
\providecommand{\url}[1]{\href{#1}{#1}}

\bibitem[{{Acu{\~n}a} {et~al.}(1995){Acu{\~n}a}, {Ogilvie}, {Baker}, {Curtis},
  {Fairfield}, \& {Mish}}]{1995SSRv...71....5A}
{Acu{\~n}a}, M.~H., {Ogilvie}, K.~W., {Baker}, D.~N., {et~al.} 1995, \ssr, 71,
  5

\bibitem[{{Baker} {et~al.}(2013){Baker}, {Brooks}, {D{\'e}moulin}, {van
  Driel-Gesztelyi}, {Green}, {Steed}, \& {Carlyle}}]{2013ApJ...778...69B}
{Baker}, D., {Brooks}, D.~H., {D{\'e}moulin}, P., {et~al.} 2013, \apj, 778, 69

\bibitem[{{Borrini} {et~al.}(1982){Borrini}, {Gosling}, {Bame}, \&
  {Feldman}}]{1982JGR....87.7370B}
{Borrini}, G., {Gosling}, J.~T., {Bame}, S.~J., \& {Feldman}, W.~C. 1982, \jgr,
  87, 7370

\bibitem[{{Cheng} \& {Ding}(2016)}]{2016ApJ...823L...4C}
{Cheng}, X., \& {Ding}, M.~D. 2016, \apjl, 823, L4

\bibitem[{{Cheung} {et~al.}(2015){Cheung}, {Boerner}, {Schrijver}, {Testa},
  {Chen}, {Peter}, \& {Malanushenko}}]{2015ApJ...807..143C}
{Cheung}, M. C.~M., {Boerner}, P., {Schrijver}, C.~J., {et~al.} 2015, \apj,
  807, 143

\bibitem[{{Cliver} {et~al.}(2003){Cliver}, {Ling}, \&
  {Richardson}}]{2003ApJ...592..574C}
{Cliver}, E.~W., {Ling}, A.~G., \& {Richardson}, I.~G. 2003, \apj, 592, 574

\bibitem[{{Davies} {et~al.}(2013){Davies}, {Perry}, {Trines}, {Harrison},
  {Lugaz}, {M{\"o}stl}, {Liu}, \& {Steed}}]{2013ApJ...777..167D}
{Davies}, J.~A., {Perry}, C.~H., {Trines}, R.~M.~G.~M., {et~al.} 2013, \apj,
  777, 167

\bibitem[{{De Pontieu} {et~al.}(2014){De Pontieu}, {Title}, {Lemen}, {Kushner},
  {Akin}, {Allard}, {Berger}, {Boerner}, {Cheung}, {Chou}, {Drake}, {Duncan},
  {Freeland}, {Heyman}, {Hoffman}, {Hurlburt}, {Lindgren}, {Mathur}, {Rehse},
  {Sabolish}, {Seguin}, {Schrijver}, {Tarbell}, {W{\"u}lser}, {Wolfson},
  {Yanari}, {Mudge}, {Nguyen-Phuc}, {Timmons}, {van Bezooijen}, {Weingrod},
  {Brookner}, {Butcher}, {Dougherty}, {Eder}, {Knagenhjelm}, {Larsen},
  {Mansir}, {Phan}, {Boyle}, {Cheimets}, {DeLuca}, {Golub}, {Gates}, {Hertz},
  {McKillop}, {Park}, {Perry}, {Podgorski}, {Reeves}, {Saar}, {Testa}, {Tian},
  {Weber}, {Dunn}, {Eccles}, {Jaeggli}, {Kankelborg}, {Mashburn}, {Pust},
  {Springer}, {Carvalho}, {Kleint}, {Marmie}, {Mazmanian}, {Pereira}, {Sawyer},
  {Strong}, {Worden}, {Carlsson}, {Hansteen}, {Leenaarts}, {Wiesmann},
  {Aloise}, {Chu}, {Bush}, {Scherrer}, {Brekke}, {Martinez-Sykora}, {Lites},
  {McIntosh}, {Uitenbroek}, {Okamoto}, {Gummin}, {Auker}, {Jerram}, {Pool}, \&
  {Waltham}}]{2014SoPh..289.2733D}
{De Pontieu}, B., {Title}, A.~M., {Lemen}, J.~R., {et~al.} 2014, \solphys, 289,
  2733

\bibitem[{{Feldman}(1992)}]{1992PhyS...46..202F}
{Feldman}, U. 1992, \physscr, 46, 202

\bibitem[{{Gloeckler} {et~al.}(1998){Gloeckler}, {Cain}, {Ipavich}, {Tums},
  {Bedini}, {Fisk}, {Zurbuchen}, {Bochsler}, {Fischer}, {Wimmer-Schweingruber},
  {Geiss}, \& {Kallenbach}}]{1998SSRv...86..497G}
{Gloeckler}, G., {Cain}, J., {Ipavich}, F.~M., {et~al.} 1998, \ssr, 86, 497

\bibitem[{{Gopalswamy}(2006)}]{2006SSRv..124..145G}
{Gopalswamy}, N. 2006, \ssr, 124, 145

\bibitem[{{Gopalswamy} {et~al.}(2013){Gopalswamy}, {M{\"a}kel{\"a}}, {Akiyama},
  {Xie}, {Yashiro}, \& {Reinard}}]{2013SoPh..284...17G}
{Gopalswamy}, N., {M{\"a}kel{\"a}}, P., {Akiyama}, S., {et~al.} 2013, \solphys,
  284, 17

\bibitem[{{Harrison}(1995)}]{1995A&A...304..585H}
{Harrison}, R.~A. 1995, \aap, 304, 585

\bibitem[{{Harrison}(2003)}]{2003AdSpR..32.2425H}
---. 2003, Advances in Space Research, 32, 2425

\bibitem[{{Laming}(2015)}]{2015LRSP...12....2L}
{Laming}, J.~M. 2015, Living Reviews in Solar Physics, 12, 2

\bibitem[{{Landi} {et~al.}(2013){Landi}, {Young}, {Dere}, {Del Zanna}, \&
  {Mason}}]{2013ApJ...763...86L}
{Landi}, E., {Young}, P.~R., {Dere}, K.~P., {Del Zanna}, G., \& {Mason}, H.~E.
  2013, \apj, 763, 86

\bibitem[{{Lemen} {et~al.}(2012){Lemen}, {Title}, {Akin}, {Boerner}, {Chou},
  {Drake}, {Duncan}, {Edwards}, {Friedlaender}, {Heyman}, {Hurlburt}, {Katz},
  {Kushner}, {Levay}, {Lindgren}, {Mathur}, {McFeaters}, {Mitchell}, {Rehse},
  {Schrijver}, {Springer}, {Stern}, {Tarbell}, {Wuelser}, {Wolfson}, {Yanari},
  {Bookbinder}, {Cheimets}, {Caldwell}, {Deluca}, {Gates}, {Golub}, {Park},
  {Podgorski}, {Bush}, {Scherrer}, {Gummin}, {Smith}, {Auker}, {Jerram},
  {Pool}, {Soufli}, {Windt}, {Beardsley}, {Clapp}, {Lang}, \&
  {Waltham}}]{2012SoPh..275...17L}
{Lemen}, J.~R., {Title}, A.~M., {Akin}, D.~J., {et~al.} 2012, \solphys, 275, 17

\bibitem[{{Lepping} {et~al.}(1995){Lepping}, {Ac{\~{u}}na}, {Burlaga},
  {Farrell}, {Slavin}, {Schatten}, {Mariani}, {Ness}, {Neubauer}, {Whang},
  {Byrnes}, {Kennon}, {Panetta}, {Scheifele}, \&
  {Worley}}]{1995SSRv...71..207L}
{Lepping}, R.~P., {Ac{\~{u}}na}, M.~H., {Burlaga}, L.~F., {et~al.} 1995, \ssr,
  71, 207

\bibitem[{{Lepri} \& {Zurbuchen}(2004)}]{2004JGRA..109.1112L}
{Lepri}, S.~T., \& {Zurbuchen}, T.~H. 2004, Journal of Geophysical Research
  (Space Physics), 109, A01112

\bibitem[{{Lepri} {et~al.}(2001){Lepri}, {Zurbuchen}, {Fisk}, {Richardson},
  {Cane}, \& {Gloeckler}}]{2001JGR...10629231L}
{Lepri}, S.~T., {Zurbuchen}, T.~H., {Fisk}, L.~A., {et~al.} 2001, \jgr, 106,
  29231

\bibitem[{{Lin} {et~al.}(2005){Lin}, {Ko}, {Sui}, {Raymond}, {Stenborg},
  {Jiang}, {Zhao}, \& {Mancuso}}]{2005ApJ...622.1251L}
{Lin}, J., {Ko}, Y.~K., {Sui}, L., {et~al.} 2005, \apj, 622, 1251

\bibitem[{{Liu} {et~al.}(2018){Liu}, {Su}, {Cheng}, {van Ballegooijen}, \&
  {Ji}}]{2018ApJ...868...59L}
{Liu}, T., {Su}, Y., {Cheng}, X., {van Ballegooijen}, A., \& {Ji}, H. 2018,
  \apj, 868, 59

\bibitem[{{Manchester} {et~al.}(2017){Manchester}, {Kilpua}, {Liu}, {Lugaz},
  {Riley}, {T{\"o}r{\"o}k}, \& {Vr{\v{s}}nak}}]{2017SSRv..212.1159M}
{Manchester}, W., {Kilpua}, E. K.~J., {Liu}, Y.~D., {et~al.} 2017, \ssr, 212,
  1159

\bibitem[{{Neugebauer} \& {Goldstein}(1997)}]{1997GMS....99..245N}
{Neugebauer}, M., \& {Goldstein}, R. 1997, Washington DC American Geophysical
  Union Geophysical Monograph Series, 99, 245

\bibitem[{{Ogilvie} {et~al.}(1995){Ogilvie}, {Chornay}, {Fritzenreiter},
  {Hunsaker}, {Keller}, {Lobell}, {Miller}, {Scudder}, {Sittler}, {Torbert},
  {Bodet}, {Needell}, {Lazarus}, {Steinberg}, {Tappan}, {Mavretic}, \&
  {Gergin}}]{1995SSRv...71...55O}
{Ogilvie}, K.~W., {Chornay}, D.~J., {Fritzenreiter}, R.~J., {et~al.} 1995,
  \ssr, 71, 55

\bibitem[{{Owens}(2018)}]{2018SoPh..293..122O}
{Owens}, M.~J. 2018, \solphys, 293, 122

\bibitem[{{Pesnell} {et~al.}(2012){Pesnell}, {Thompson}, \&
  {Chamberlin}}]{2012SoPh..275....3P}
{Pesnell}, W.~D., {Thompson}, B.~J., \& {Chamberlin}, P.~C. 2012, \solphys,
  275, 3

\bibitem[{{Priest} \& {Forbes}(2002)}]{2002A&ARv..10..313P}
{Priest}, E.~R., \& {Forbes}, T.~G. 2002, \aapr, 10, 313

\bibitem[{{Reinard}(2008)}]{2008ApJ...682.1289R}
{Reinard}, A.~A. 2008, \apj, 682, 1289

\bibitem[{{Richardson} \& {Cane}(2004)}]{2004JGRA..109.9104R}
{Richardson}, I.~G., \& {Cane}, H.~V. 2004, Journal of Geophysical Research
  (Space Physics), 109, A09104

\bibitem[{{Schou} {et~al.}(2012){Schou}, {Scherrer}, {Bush}, {Wachter},
  {Couvidat}, {Rabello-Soares}, {Bogart}, {Hoeksema}, {Liu}, {Duvall}, {Akin},
  {Allard}, {Miles}, {Rairden}, {Shine}, {Tarbell}, {Title}, {Wolfson},
  {Elmore}, {Norton}, \& {Tomczyk}}]{2012SoPh..275..229S}
{Schou}, J., {Scherrer}, P.~H., {Bush}, R.~I., {et~al.} 2012, \solphys, 275,
  229

\bibitem[{{SONG} {et~al.}(2015){SONG}, {CHEN}, {ZHANG}, {CHENG}, {Wang}, {HU},
  {LI}, \& {WANG}}]{2015ApJ...808L..15S}
{SONG}, H.~Q., {CHEN}, Y., {ZHANG}, J., {et~al.} 2015, \apjl, 808, L15

\bibitem[{{Song} {et~al.}(2016){Song}, {Zhong}, {Chen}, {Zhang}, {Cheng},
  {Zhao}, {Hu}, \& {Li}}]{2016ApJS..224...27S}
{Song}, H.~Q., {Zhong}, Z., {Chen}, Y., {et~al.} 2016, \apjs, 224, 27

\bibitem[{{Stone} {et~al.}(1998){Stone}, {Frandsen}, {Mewaldt}, {Christian},
  {Margolies}, {Ormes}, \& {Snow}}]{1998SSRv...86....1S}
{Stone}, E.~C., {Frandsen}, A.~M., {Mewaldt}, R.~A., {et~al.} 1998, \ssr, 86, 1

\bibitem[{{Su} {et~al.}(2018){Su}, {Veronig}, {Hannah}, {Cheung}, {Dennis},
  {Holman}, {Gan}, \& {Li}}]{2018ApJ...856L..17S}
{Su}, Y., {Veronig}, A.~M., {Hannah}, I.~G., {et~al.} 2018, \apjl, 856, L17

\bibitem[{{Su} {et~al.}(2013){Su}, {Veronig}, {Holman}, {Dennis}, {Wang},
  {Temmer}, \& {Gan}}]{2013NatPh...9..489S}
{Su}, Y., {Veronig}, A.~M., {Holman}, G.~D., {et~al.} 2013, Nature Physics, 9,
  489

\bibitem[{{Tian} {et~al.}(2013){Tian}, {Tomczyk}, {McIntosh}, {Bethge}, {de
  Toma}, \& {Gibson}}]{2013SoPh..288..637T}
{Tian}, H., {Tomczyk}, S., {McIntosh}, S.~W., {et~al.} 2013, \solphys, 288, 637

\bibitem[{{Tian} {et~al.}(2015){Tian}, {Young}, {Reeves}, {Chen}, {Liu}, \&
  {McKillop}}]{2015ApJ...811..139T}
{Tian}, H., {Young}, P.~R., {Reeves}, K.~K., {et~al.} 2015, \apj, 811, 139

\bibitem[{{Yang} {et~al.}(2019){Yang}, {Zhang}, {Song}, {Bi}, \&
  {Li}}]{2019ApJ...878...38Y}
{Yang}, S., {Zhang}, J., {Song}, Q., {Bi}, Y., \& {Li}, T. 2019, \apj, 878, 38

\bibitem[{{Young} {et~al.}(2013){Young}, {Doschek}, {Warren}, \&
  {Hara}}]{2013ApJ...766..127Y}
{Young}, P.~R., {Doschek}, G.~A., {Warren}, H.~P., \& {Hara}, H. 2013, \apj,
  766, 127

\bibitem[{{Zhang} {et~al.}(2012){Zhang}, {Cheng}, \&
  {Ding}}]{2012NatCo...3..747Z}
{Zhang}, J., {Cheng}, X., \& {Ding}, M.-D. 2012, Nature Communications, 3, 747

\bibitem[{{Zhang} {et~al.}(2001){Zhang}, {Dere}, {Howard}, {Kundu}, \&
  {White}}]{2001ApJ...559..452Z}
{Zhang}, J., {Dere}, K.~P., {Howard}, R.~A., {Kundu}, M.~R., \& {White}, S.~M.
  2001, \apj, 559, 452

\bibitem[{{Zhu} {et~al.}(2016){Zhu}, {Wang}, {Du}, \&
  {He}}]{2016ApJ...826...51Z}
{Zhu}, X., {Wang}, H., {Du}, Z., \& {He}, H. 2016, \apj, 826, 51

\bibitem[{{Zhu} {et~al.}(2013){Zhu}, {Wang}, {Du}, \&
  {Fan}}]{2013ApJ...768..119Z}
{Zhu}, X.~S., {Wang}, H.~N., {Du}, Z.~L., \& {Fan}, Y.~L. 2013, \apj, 768, 119

\end{thebibliography}



\end{document}